# Implementing a Concept Network Model

Sarah H. Solomon[1], John D. Medaglia[2,3], and Sharon L. Thompson-Schill[1]

[1] Department of Psychology, University of Pennsylvania
[2] Department of Psychology, Drexel University
[3] Department of Neurology, Perelman School of Medicine, University of Pennsylvania

May 22, 2018

## Abstract

The same concept can mean different things or be instantiated in different forms depending on context, suggesting a degree of flexibility within the conceptual system. We propose that a compositional network model can be used to capture and predict this flexibility. We modeled individual concepts (e.g., BANANA, BOTTLE) as graph-theoretical networks, in which properties (e.g., YELLOW, SWEET) were represented as nodes and their associations as edges. In this framework, networks capture the within-concept statistics that reflect how properties correlate with each other across instances of a concept. We ran a classification analysis using graph eigendecomposition to validate these models, and find that these models can successfully discriminate between object concepts. We then computed formal measures from these concept networks and explored their relationship to conceptual structure. We find that diversity coefficients and core-periphery structure can be interpreted as network-based measures of conceptual flexibility and stability, respectively. These results support the feasibility of a concept network framework and highlight its ability to formally capture important characteristics of the conceptual system.

# Introduction

The APPLE information evoked by "apple pie" is considerably different from that evoked by "apple picking": the former is soft, warm, and wedge-shaped, whereas the latter is firm, cool, and spherical. If you scour your conceptual space for APPLE information, you will uncover the knowledge that apples can be red, green, yellow, or brown when old; that they can be sweet or tart; that they are crunchy when fresh and soft when baked; that they are naturally round but can be cut into slices; that they are firm, but mushy if blended; that they can be found in bowls, in jars, and on trees. Despite the complexity of this conceptual knowledge, we can generate an appropriate APPLE instance, with the appropriate features, based on the context we are in at the time. In other words, the multi-faceted APPLE concept can be flexibly adjusted in order to enable a near-infinite number of specific and appropriate APPLE exemplars.

How is conceptual knowledge structured such that this generative and essential flexibility is possible? In particular, we are interested in the structure of individual concepts (e.g., APPLE, SNOW), rather than the structure of super-ordinate categories (e.g., FRUIT, TOOLS) or the structure of semantic space more broadly. This latter pursuit — the modeling of semantic space — has already been approached from various theoretical orientations and methodologies. In "compositional" cognitive theories, the meaning of a concept can be decomposed into features and their relationships with each other (e.g., Smith et al, 1974, McRae et al., 1997; Tyler & Moss, 2001). This compositionality can be incorporated into the associated computational models: researchers primarily use various forms of feature-based connectionist models, in which concepts are represented as patterns of activation over features, to simulate semantic behavior (e.g., Cree et al., 1999; 2006; Randall et al., 2004). In "relational" frameworks, concepts are defined in terms of how they relate to other concepts in semantic or lexical space (e.g., Landauer & Dumais, 1997). Researchers within this framework can use word co-occurrence or association statistics to create large semantic networks, which can be analyzed using a rich set of network science tools (e.g., Steyvers & Tenenbaum, 2005; Van Rensbergen et al., 2015; De Deyne et al, 2016).

The compositional approach to conceptual knowledge generally represents individual concepts as vectors of features. These features can span a range of information-types (e.g., visual, functional, encyclopedic), consistent with a distributed account of conceptual knowledge. The "conceptual structure" account (Tyler & Moss, 2001) represents concepts as binary vectors indicating the presence of absence of features, and argues that broad semantic domains (e.g., ANIMALS, TOOLS) differ in their characteristic properties and in their patterns of property-correlations (e.g., HAS-WINGS and FLIES tend to co-occur within the ANIMAL domain). The "feature-correlation" account (e.g., McRae et al., 1997; 1999; McRae, 2004) tweaks this model by empirically deriving conceptual property statistics and by implementing this framework in a type of connectionist model called an attractor network: property statistics characterize the structure of conceptual space, and the model can leverage these statistics to settle on an appropriate conceptual representation given the current inputs (Cree et al., 1999; Cree et al., 2006). These models contain a dynamic component in that the attractor networks reveal how word comprehension may unfold over time, and is "flexible" in the sense that the model may follow varied trajectories through conceptual space in order to settle on a specific concept's representation. However, the ability of this approach to capture conceptual flexibility — in the sense described above — is not fully fleshed out. A feature-based framework is valuable because the elements that can be adjusted during conceptual processing are explicitly modeled, but a different set of tools may be helpful in the pursuit of capturing how flexibility might emerge out of this conceptual structure. Our initial steps to develop such tools are reported here.

Another way to model conceptual knowledge is to capture statistical relations *between* words or phrases in language. This approach is "relational", rather than compositional, because a concept's meaning is represented in terms of its relations to other concepts, rather than assuming any kind of internal conceptual structure. Word co-occurrence statistics can be extracted from text corpora and have been used to create probabilistic models of word meanings (Griffiths et al., 2007), to represent semantic similarity (Landauer & Dumais, 1997), and to characterize the structure of the entire lexicon (e.g., WordNet; Miller & Fellbaum, 2007). In a similar approach, word association data can be used to capture and analyze the structure of semantic space (Steyvers & Tenenbaum, 2005; Van Rensbergen et al., 2015; De Deyne et al., 2016). These data are generally modeled as networks, which can be analyzed in formal ways.

The use of networks to model semantic knowledge has a well-established history. The early "semantic network" models (Collins & Quillian, 1969; Collins & Loftus, 1975) represent concepts as nodes in a network; links between these nodes signify associations in semantic memory. These networks capture the extent to which concepts are related to other concepts and features, and can model the putatively hierarchical nature of conceptual knowledge. Though these models are "network-based", they are so in a rather informal way. On the other hand, network science, a mathematical descendent of graph theory, has developed a rich set of tools to study networks in a formal, quantitative framework (Barabási, 2016). There has been a recent surge of research applying network science to neural and linguistic data, providing new insights into these complex systems along with new tools and measures we can use to study them.

Formal networks are composed of units (i.e., "nodes") and the links between them (i.e., "edges"). Current network science approaches to semantic and lexical knowledge use nodes to represent individual words, and edges to represent their co-occurrence or association statistics. Once modeled in this way, aspects of network structure can be quantitatively analyzed and relationships between network structure and other phenomena can be explored. For example, it has been suggested that human language exhibits small-world properties (Steyvers & Tenenbaum, 2005; i Cancho & Sole, 2001), and that semantic networks exhibit an "assortative" structure, meaning that semantic nodes tend to have connections to other semantic nodes with similar characteristics (e.g., valence, arousal, concreteness; Van Rensbergen et al., 2015). A spreading activation model applied to these word-association networks makes accurate predictions of weak similarity judgments (De Deyne et al., 2016). Further, Steyvers & Tenenbaum (2005) report that a word's degree (i.e., how many links it has to other nodes) predicts both age of acquisition and reaction times on lexical decision tasks. The application of network science tools to semantic data enables researchers to explore higher-level structure in semantic space, and to use these structural characteristics to predict aspects of semantic and lexical processing. However, this approach does not examine the internal structure of concepts: individual concepts are characterized in terms of their statistical co-occurrences with other concepts in language, and not in terms of their unique, internal content. That is, these language-based approaches are non-compositional, and we argue that compositionality is a key aspect of flexible conceptual models. Modeling a concept's internal structure — along with its features and the ways those features interact — enables us to flexibly adjust which features are included, and to what degree, in a given conceptual instance.

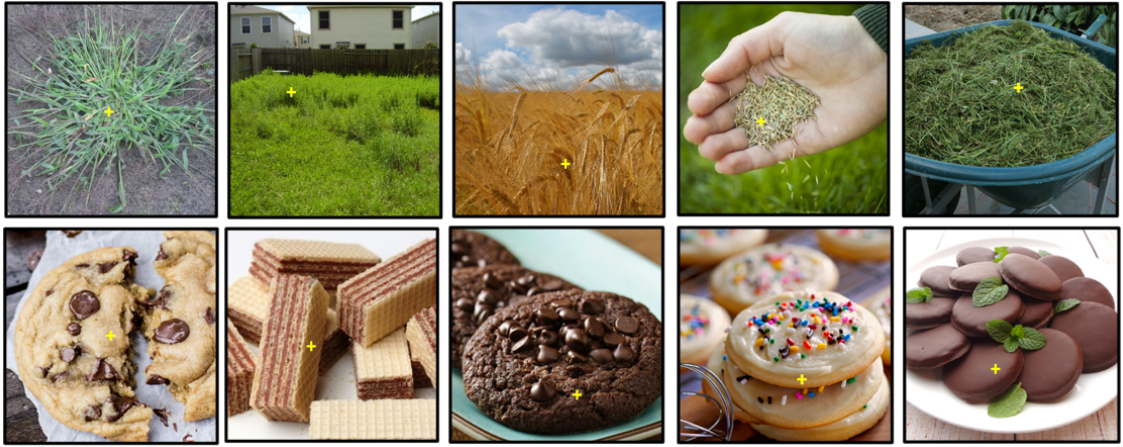

**Figure 1: Example images used to generate test data in classification analysis.** Test data used in the classification analysis were generated from participants who made property judgments on images of conceptual exemplars. Yellow cross indicates object to be considered. Example images for grass (top) and cookie (bottom).

We believe that a compositional conceptual framework paired with network science techniques provides a platform on which to model conceptual flexibility. Unlike previous network approaches, we use networks to model individual concepts, rather than the semantic system as a whole. In this case, the nodes in each concept's network represent individual features, and the edges of the network (i.e., the links between the nodes) represent the statistical relationship between features within that concept. That is, edges capture the extent to which certain properties tend to covary with each other within a concept. The creation of such networks thus depends on our ability to calculate within-concept statistics. These statistics provide the scaffolding to build our networks, and also reveal how a concept's information may be appropriately adjusted to form valid, yet varied, instances of that concept. Our goal is to show that creation of such networks is possible, and that they can be used to capture conceptual flexibility, and other phenomena, in a formal way.

We chose 15 basic-level concepts (e.g., CHOCOLATE, TABLE, GRASS, KNIFE) and defined a list of within-concept states for each one (e.g., DARK

CHOCOLATE, WHITE CHOCOLATE, CHOCOLATE SYRUP, CHOCOLATE CHIPS) using a large sample of participants on Amazon Mechanical Turk (AMT). We then compiled a large set of conceptual properties that could apply to any of the 15 concepts (e.g., BROWN, GREEN, WOODEN, METAL, SHARP, SWEET). A final sample of AMT participants then reported which of the properties corresponded to each of the specific concept-states. This let us know which properties applied to dark chocolate (e.g., BLACK, BITTER) and which applied to white chocolate (e.g., WHITE, SWEET). These data enable us to calculate the within-concept statistics necessary to construct our networks. Each of the properties corresponds to a vector that denotes whether that property is present or absent in each of the concept-states, and we can correlate these property vectors with each other to determine the extent to which each property covaries with every other property within that specific concept. These correlation values are encoded as the edges in the concept's network.

Once these networks were constructed, we had two objectives: First, we intended to determine whether or not these networks contain concept-specific information. We thus ran a classification analysis over these networks to confirm that within-concept statistics can be used to discriminate between concepts. We performed eigendecomposition on our concept networks in a classification analysis, which provided a measure of the extent to which a vector is consistent with an underlying network structure (e.g., Medaglia et al., 2017, Huang et al., in press). In our case, our test data were vectors of properties generated from photographs of individual concept exemplars (Fig. 1). Second — and most importantly — we intended to extract useful and interpretable measures from our concept networks. In particular, we aimed to quantify conceptual flexibility.

Many networks by their very nature permit flexibility, because a single network can support different states, each characterized by different patterns of activation across nodes. A node's contribution to network flexibility can be understood in terms of its position in the context of the larger network. Most natural systems exhibit "small-world" network structure (Bassett & Bullmore, 2017), which means that there are clusters of nodes in a network with strong connections between them (Amaral et al., 2000). These are called "modules", and nodes can interact with these modules in different ways. Some nodes may have links that are highly distributed across the modules in a network, whereas other nodes may have links only in one module. Each node in a network can be assigned a diversity coefficient, a version of the participation coefficient calculated using normalized Shannon entropy, which reflects this tendency. We interpreted network diversity as a likely candidate for a formal flexibility measure, and pooled the diversity coefficients across nodes in a concept's network to quantify that concept's flexibility. We calculated this version of flexibility for our 15 basic-level concepts, and predicted that it would correlate with a measure of "semantic diversity" calculated separately using word co-occurrence statistics (SemD; Hoffman et al., 2013). This would suggest that network-based measures can successfully be used to quantify flexibility in a compositional conceptual framework.

Another phenomenon of interest to cognitive scientists is the distinction between context-independent and context-dependent conceptual properties (Barsalou, 1982). Context-independent properties are those that are automatically activated for a concept in all contexts, and are sometimes referred to as "core" properties. On the other hand, context-dependent properties are those that are only activated when the context renders them relevant. Concepts are composed of both kinds of properties, such that some properties are stable and are activated across all instances, and some are more variable and are only activated some of the time. The distinction between context-independent and –dependent properties has been suggested in reaction time differences in property-verification tasks (Barsalou, 1982), but the classification of a property as one type or the other has been decided upon by the experimenter rather than being calculated in a quantitative way. One of our goals was to use our concept networks to extract this information; that is, whether the structure of each concept does in fact include such a core.

Network science provides techniques for assessing this core-periphery structure (Borgatti & Everett, 2000; Bassett et al., 2013). In network terms, a core is a set of nodes that are densely interconnected and therefore often co-activated, whereas the periphery consists of nodes with sparser connections. A measure can be extracted that represents the extent to which a given network has a core-periphery structure; some networks might have more prominent cores than others. We hypothesize that this construct of core-periphery structure can provide a way to formally capture the notion of context-dependent and context-independent conceptual properties. In other words, perhaps a certain concept has a large set of context-independent properties that are consistently activated across a large range of contexts: this concept's network might have a strong core-periphery structure. It also seems reasonable to suggest that those concepts with a stronger core might be less flexible in the ways described above. If we interpret a core as a set of properties whose activation patterns are stable across contexts, then there is less room for variability in the expression of those properties, and therefore less flexibility overall. On the other hand, more flexible concepts might have a weaker core-periphery structure, reflecting the more variable patterns of property activations. We therefore predict a negative relationship between our network measures of flexibility and core-periphery structure.

## Methods

### General Methods

**Network Construction** In order to create our networks we first had to define our nodes. Since our nodes represent individual conceptual properties, we compiled a list of properties that could be applied to all of our target concepts. Participants were recruited from Amazon Mechanical Turk (AMT) and were asked to list all of the properties that must be true or can be true for each concept. It was emphasized that the properties do not have to be true of all types of the concept. Participants were required to report at least 10 properties per concept, but there was no limit on the number of responses they could provide. Once these data were collected, we organized the data as follows. For each concept, we collapsed across different forms of the same property (e.g., "sugar", "sugary", "tastes sugary"), and removed responses that were too general (e.g., "taste", "color"). For each concept, we only included properties that were given by more than one participant. We then combined properties across all concepts to create our final list of *N* properties that will be represented as nodes in our concept networks.

The same AMT participants that provided conceptual properties also provided concept-states for each of the target concepts. For each concept, participants were asked to think about that object and all the different kinds, forms, types, or states in which that object can be found. Participants were required to make at least five responses, and could make up to 15 responses. For each concept, we removed responses that we considered properties rather than types (e.g., "sweet chocolate"), and responses that were too specific (e.g., "Chiquita banana"). We only included responses that were given by more than one participant, resulting in a set of *K* concept-states for each concept.

A separate set of AMT participants was presented with one concept-state of each of the target concepts in random order (e.g., "dark chocolate", "frozen banana") and was asked to select the properties that are true of that specific concept-state. The full list of *N* properties was displayed in a multiple-choice format. For each concept-state, responses were combined across participants and represented in a binary fashion. A property was only considered "true" for a concept-state if more than one participant made that response. At this point, each concept's data included a set of *K* concept-states, each of which corresponds to a *N*-length vector that indicates the presence or absence of each property. We could also view these data as a set of *N* conceptual properties, each of which corresponded to a *K*-length vector that indicates its presence or absence in each of the concept-states.

For each concept, we excluded properties that were not present in any of the concept-states, resulting in a smaller set of $N_c$ properties. We created a network by correlating the $N_c$ binary property-vectors with each other to create a $N_c$ x $N_c$ symmetrical, weighted correlation matrix. The diagonal was set to 0, and these networks were filtered using the triangulation filtering method (Massara et al., 2016; Tumminello et al., 2005; Kenett et al., 2014). This filtering approach generates a simpler subgraph that maximizes information content while reducing the influence of noise. This method is appropriate for graphs where edges are defined as correlations between nodes, as is the case here. No parameter fitting is required. These final, filtered concept networks were then analyzed using standard network science methods.

**Classification Analysis** If our concept network models capture concept-specific information, the networks should be able to successfully discriminate between new concept exemplars. Exemplar data were generated from sets of photographs for each concept; all concept-states were represented. AMT participants were shown one image per concept, were asked to imagine interacting with this object in the real world, and to consider what properties it has. The full list of *N* properties was displayed in multiple-choice format, and participants were asked to select the properties that they believed applied to the object in the image. Individual participants' responses to each concept-state were represented as *N*-length property vectors and were used as test data in the classification analysis.

By performing eigendecomposition on each adjacency matrix (i.e., concept network) we can assess the extent to which a vector is expected given an underlying network structure (e.g., Medaglia et al., 2017; Huang et al., in press). For each adjacency matrix $A$, $V$ is the set of $N_c$ eigenvectors, ordered by eigenvalue. $M$ is the number of ordered eigenvectors to include in analysis, and designates a subset of $V$. For each eigenvector $v$, we find the dot product with signal vector $x$, which gives us the projection of $x$ on that dimension in the eigenspace of $A$. That is, it gives us an "alignment" value for that particular signal and that particular eigenvector. We can include all eigenvectors in $M$ by taking the sum of squares of the dot products for each eigenvector. The alignment value for each signal is defined as:

$$\tilde{x} = \sum_{i=1}^{M}(v_i \cdot x)^2 \qquad (1)$$

where $x$ is a property vector, $M$ is the number of eigenvectors to include in alignment (sorted by eigenvalue), $v_i$ is one of $M$ eigenvectors of the adjacency matrix, and $\tilde{x}$ is the scalar alignment value for signal $x$ with adjacency matrix $A$, given the eigenvectors 1-$M$. In our case, signal $x$ is a property vector corresponding to a particular exemplar image (e.g., Fig. 1), which we align with each of the concept networks. Each exemplar was restricted to the properties included in each concept model before transformation; that is, exemplar data ($x$) were reduced to $N_c$–length vectors. The concept network that resulted in the highest alignment value ($\tilde{x}$) was taken as the "guess" of the classifier; each exemplar was either classified correctly (1), or incorrectly (0). We averaged these data across all exemplars to calculate the average classifier accuracy.

To calculate a baseline measure of classification accuracy, we created traditional vector models for each concept. These models were similar to those used elsewhere in the literature (Tyler & Moss, 2001; McRae et al. 1997; 1999; 2004). For each concept, we averaged the $K$ concept-state vectors resulting in an $N_c$-length vector containing mean property strength values. Each concept's traditional vector model and network model contained the same conceptual properties. We ran a separate classification analysis using these traditional models and a correlational classifier. Each exemplar property-vector was correlated with each of the traditional concept vector models; the concept model that resulted in the highest correlation value was taken as the guess of the classifier. We calculated average measures of classifier performance using the same methods described above, and also calculated classification accuracy within each concept.

**Network Analysis** We extracted network metrics from our concept networks using the Brain Connectivity Toolbox (Rubinov & Sporns, 2010). The set of nodes in each network is designated as $N$, and $n$ is the number of nodes. The set of links is $L$, and $l$ is the number of links. The existence of a link between nodes $(i,j)$ is captured in $a_{ij}$: $a_{ij} = 1$ if a link is present and $a_{ij} = 0$ if a link is absent. The weight of a link is represented as $w_{ij}$, and is normalized such that $0 \leq w_{ij} \leq 1$. $l^w$ is the sum of all weights in the network. The network metrics we extracted included node strength, node degree, modularity ($Q$), core-periphery structure, and diversity coefficients (Fig. 2).

Nodes within a network differ in the number and strength of their connections to other nodes. Node degree ($k$) is the number of connections that each node has with other nodes in the network (Eq. 2; Rubinov & Sporns, 2010). In weighted (i.e., non-binary) networks, node strength ($k^w$) is calculated by summing the weights of the connections with other nodes (Eq. 3; Rubinov & Sporns, 2010). We separately averaged node strength and node degree within each network to obtain mean strength and degree measures for each concept network. We also counted the number of edges in each network overall.

$$k_i = \sum_{j \in N} a_{ij} \tag{2}$$

$$k_i^w = \sum_{j \in N} w_{ij} \tag{3}$$

Modularity ($Q$) is a metric that describes a network's community structure. We can attempt to partition a weighted network into sets of non-overlapping nodes (i.e., modules) such that within-module connections are maximized and between-module connections are minimized. Some networks exhibit more of a modular structure than others; $Q^w$ is a quantitative measure of modularity for each weighted network (Eq. 4; Rubinov & Sporns, 2010), which is defined as

$$Q^w = \frac{1}{l^w} \sum_{i,j \in N} \left[ w_{ij} - \frac{k_i^w k_j^w}{l^w} \right] \delta_{m_i, m_j} \tag{4}$$

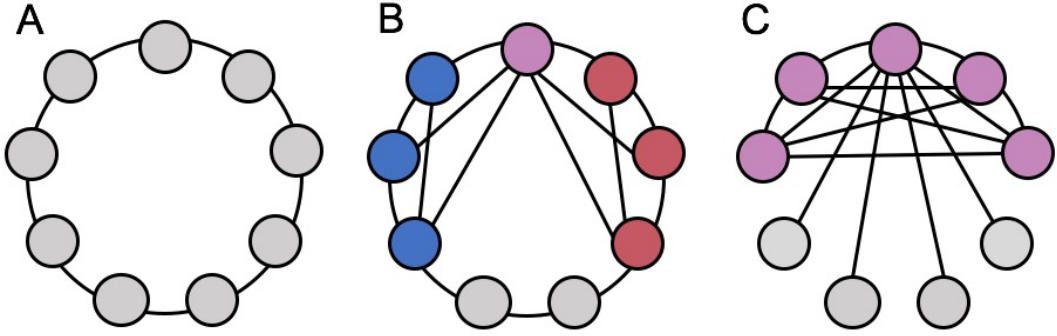

**Figure 2: Schematics of network structure.** (A) Low-modularity network that contains nodes with equal degree. (B) High-modularity network with nodes in either module 1 (red) or module 2 (blue). One node (purple) participates in both modules; this is a high-diversity node. (C) Network with a strong core-periphery structure; some nodes comprise a densely connected core (purple) and others a weakly connected periphery (grey).

where $\delta_{m_i,m_j} = 1$ if nodes $i,j$ are in the same module ($m$), $w_{ij}$ is the specific strength between nodes $i,j$, and $\frac{k_i^w k_j^w}{l^w}$ scales $w_{ij}$ by the total strengths of nodes $i,j$ across the network. Given a network's community structure, we can observe how individual nodes participate with each of the modules in the set of modules ($M$): Nodes may have connections to many different modules, or have very few such connections. The diversity coefficient ($h_i^{\pm}$) is a measure ascribed to individual nodes that reflects the diversity of connections that each node has to modules in the network. This is a version of the participation coefficient, and is calculated using normalized Shannon entropy; we have previously used entropy to model property flexibility, and so predicted that diversity would be a good candidate for a network-based measure of conceptual flexibility. The diversity coefficient (Eq. 5; Rubinov & Sporns, 2011) for each node is defined as

$$h_i^{\pm} = -\frac{1}{\log m} \sum_{u \in M} p_i^{\pm}(u) \log p_i^{\pm}(u), \qquad (5)$$

where $p_i^{\pm}(u) = \frac{s_i^{\pm}(u)}{s_i^{\pm}}$, $s_i^{\pm}(u)$ is the strength of node $i$ within module $u$, and $m$ is the number of modules in modularity partition $M$. We averaged diversity coefficients across nodes in a network to obtain a mean measure of diversity for each concept network.

Core-periphery structure is another way to describe the structure of a network. Here, we attempt to partition a network into two non-overlapping sets of nodes such that connections within one set are maximized (i.e., the "core") and connections in the other are minimized (i.e., the "periphery"). Core-periphery fit ($Q_C$) is a quantitative measure of how well each network can be partitioned in this way (Eq. 6), and can be defined as

$$Q_C = \frac{1}{v_C} \left( \sum_{i,j \in C_c}(w_{ij} - \gamma_C \bar{w}) - \sum_{i,j \in C_p}(w_{ij} - \gamma_C \bar{w}) \right) \tag{6}$$

where $C_c$ is the set of all nodes in the core, $C_p$ is the set of nodes in the periphery, $\bar{w}$ is the average edge weight, $\gamma_C$ is a parameter controlling the size of the core, and $v_C$ is a normalization constant (Rubinov et al., 2015).

**Methods: Set 1**

The 5 concepts used in Set 1 were CHOCOLATE, BANANA, BOTTLE, TABLE, and PAPER. AMT participants (*N*=66) provided general properties for each concept along with concept-states. Another group of AMT participants (*N*=198) made property judgments on specific concept-states, and another group of AMT participants (*N*=60) generated test data for the classification analysis by making property judgments on individual images. The final property list included 129 properties. The number of states for each concept were as follows: chocolate=14, banana=15, bottle=11, table=14, paper=20. The full list of states can be seen in the Appendix. In the classification analysis, test data comprised a total of 300 property-vectors, with 60 exemplars/concept.

**Methods: Set 2**

The 10 concepts used in Set 2 were KEY, PUMPKIN, GRASS, COOKIE, PICKLE, KNIFE, PILLOW, WOOD, PHONE, and CAR. AMT participants (*N*=60) provided general properties for each concept along with concept-states. Another group of AMT participants (*N*=108) made property judgments on specific concept-states, and another group of AMT participants (*N*=30) generated test data for the classification analysis by making property judgments on individual images. The final property list included 276 properties. The number of states for each concept were as follows: key=19, pumpkin=18, grass=16, cookie=22, pickle=17, knife=15, pillow=16, wood=22, phone=16, car=20. The full list of states can be seen in the Appendix. In the classification analysis, test data comprised 300 property-vectors, with 30 exemplars/concept.

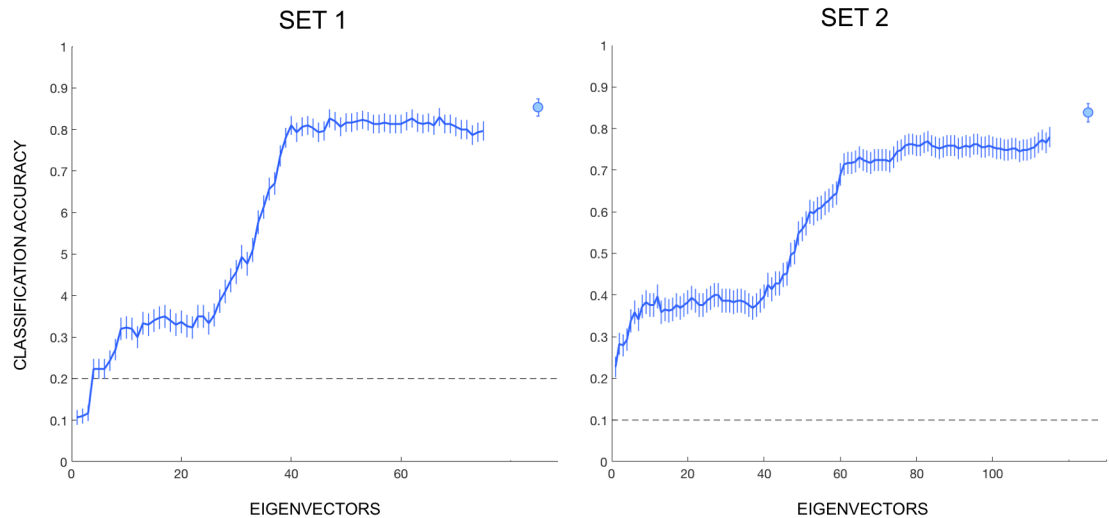

**Figure 3: Classification results.** We ran a range of classification analyses using different numbers of eigen-dimensions from our concept networks. Classification was successful using ≥ 7 dimensions in Set 1, and ≥1 dimension in Set 2. Classification performance increased as more dimensions were added, such that performance of the network-models approached performance of the vector-based models (single data points). The sharp increase in performance in both sets is driven by eigenvectors with eigenvalues of 0, suggesting that contribution of individual features, suggesting that the presence

## Results

### Classification Results

In order to determine whether our concept networks contained concept-specific information, we ran a classification analysis using eigendecomposition for both Set 1 and Set 2. We ran multiple analyses using different ranges of eigenvectors, which were sorted by eigenvalue (positive to negative). We started by only using the first eigenvector in each of the concept networks and determined whether this dimension alone could be used to classify the property vector. One dimension was enough to classify exemplars in Set 2 (Mean Accuracy=0.27; SE=0.03; Chance=0.10) but not Set 1 (Mean Accuracy=0.11; SE=0.02; Chance=0.20).

However, increasing the number of dimensions improved classification performance for both sets (Fig. 3): for example, classification performance is significantly above chance when only 10 dimensions are used in Set 1 (Mean Accuracy=0.38; SE=0.03; Chance=0.10) and Set 2 (Mean Accuracy=0.38; SE=0.03; Chance=0.20). As more dimensions were included in the analysis, classification performance significantly increases and approaches the performance of the vector-based classifier, which was successful at classifying exemplars in Set 1 (Mean Accuracy=0.85; SD=0.06; Chance=.20) and Set 2 (Mean Accuracy=0.84; SD=0.10; Chance=0.10).

|  | SemS | SemD | Modularity | Core Fit | Diversity | Strength | Degree | Edges | Vector Acc. |
|---|---|---|---|---|---|---|---|---|---|
| SemS |  |  |  |  |  |  |  |  |  |
| SemD | -0.96** |  |  |  |  |  |  |  |  |
| Modularity | -0.19 | 0.22 |  |  |  |  |  |  |  |
| Core Fit | 0.54* | -0.50 | -0.55* |  |  |  |  |  |  |
| Diversity | -0.60* | 0.56* | -0.11 | -0.61* |  |  |  |  |  |
| Strength | -0.37 | 0.37 | 0.75** | -0.64* | 0.08 |  |  |  |  |
| Degree | -0.52* | 0.50 | 0.75** | -0.88** | 0.42 | 0.84** |  |  |  |
| Edges | -0.14 | 0.11 | 0.44 | -0.64* | 0.30 | 0.71** | 0.67** |  |  |
| Vector Acc. | 0.41 | -0.42 | -0.62* | 0.56* | -0.27 | -0.58* | -0.66** | -0.70** |  |

**Table 2: Correlation results.** We analyzed relationships between cognitive variables (SemS, SemD), network variables (modularity, core fit, mean diversity, mean strength, mean degree, number of edges) and vector-based classification results. *: $p<0.05$, **: $p<0.01$.

The middle range on the x-axis of Fig. 3 contains eigenvectors with eigenvalues of 0: this is a special case in which a single property is weighted as 1. When one of these eigenvectors is multiplied by a signal (i.e., property vector) that includes that particular property, the alignment value is driven by the presence of that one property. In other words, the dramatic increase in classification performance in both Set 1 and Set 2 is driven by the successive contributions of individual features moving from left to right, suggesting that the presence or absence of individual features is highly informative for discriminating *between* concepts. Nevertheless, the significant classification performance using eigenvectors representing multiple features does suggest that our concept networks contain concept-specific information, motivating us to look *within* a concept for structural elements that relate to conceptual flexibility. It is this main goal that we pursue in the subsequent analyses.

**Network Measures of Conceptual Structure**

Networks across the two sets differed in node assignments, since they were constructed using different properties. However, once classification and network measures were extracted, we could pool the concepts together (*N*=15) and examine relationships between these network-related measures and other variables of interest.

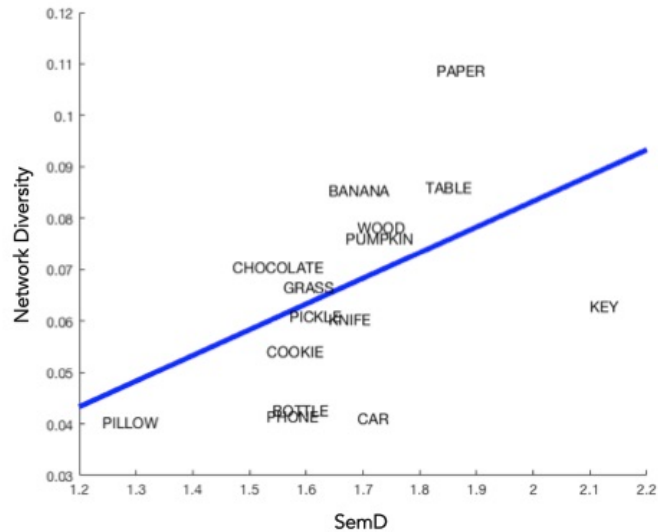

**Figure 4: A network-based measure of conceptual flexibility.** Semantic diversity measures calculated using word co-occurrence statistics (Hoffman et al., 2013) predict network diversity across 15 concepts; network diversity is mean diversity coefficient across nodes.

We extracted network measures from the full concept networks and explored how they relate to cognitive measures of conceptual flexibility and stability. Hoffman et al. (2013) use word co-occurrence statistics to quantify the context-dependent variations in word meanings found in language. The authors provide a measure of semantic diversity (SemD) that captures this variability, and we extracted SemD values for our 15 concepts. We also extracted their reported mean cosine similarity of a word's contexts and used this as a measure of semantic stability (which we refer to as SemS). As expected, SemD negatively correlated with SemS across our 15 concepts ($r(15)$=-0.96, $p$=<0.0001). The correlations between all measures of interest are shown in Table 2.

One of our primary goals was to extract a network measure that reflects conceptual flexibility. We used SemD (Hoffman et al., 2013) as a benchmark for conceptual flexibility and determined whether our hypothesized network measures of flexibility correlated with SemD across our 15 concepts. *A priori*, we hypothesized that the mean diversity (i.e., the average of a concept network's diversity coefficients across nodes) could reflect conceptual flexibility. This network measure captures the extent to which properties within a concept associate with different modules, or property clusters. Another possible candidate measure was network modularity, which reflects the extent to which a concept's network can be partitioned into separate property clusters. Network modularity was not significantly predicted by either SemD ($r(15)=0.22$, $p>0.4$) or SemS ($r(15)=-0.19$, $p>0.5$). On the other hand, mean diversity was positively predicted by SemD ($r(15)=0.56$, $p=0.03$; Fig. 4) and negatively predicted by SemS ($r(15)=-0.60$, $p=0.02$). Mean diversity was not significantly predicted by mean node strength ($r(15)=-0.08$, $p>0.7$), mean node degree ($r(15)=0.42$, $p=0.12$), or total number of network edges ($r(15)=0.3$, $p>0.2$). These results suggest that the network measure of mean diversity is a strong candidate for a quantitative measure of conceptual flexibility.

We also assessed the core-periphery structure for each concept network, which determines how well a network can be divided into a densely connected core and a sparsely connected periphery. If the core of a concept network corresponds to the notion of a context-independent conceptual "core", we would expect that more stable (i.e., less flexible) concepts would have networks with a stronger core-periphery structure. Consistent with this prediction, core fit was positively predicted by SemS ($r(15)=0.54$, $p=0.04$), though the relationship with SemD was only marginally significant ($r(15)=-0.50$, $p=0.06$). Furthermore, mean diversity and core fit were negatively correlated ($r(15)=-0.61$, $p=0.02$), suggesting that these measures may be used to capture conceptual flexibility and stability, respectively. We also found that classification accuracy using the traditional vector model was positively correlated with core fit ($r(15)=0.56$, $p=0.03$). This suggests that standard cognitive models perform better on more stable concepts, highlighting the need for a model that can adequately capture conceptual flexibility.

## Discussion

Here our goal was to model basic-level concepts using graph-theoretical networks within a compositional conceptual framework. We argue that the within-concept statistics encoded in these models capture useful, concept-specific information. Using standard network science tools, we further reveal the usefulness of these models by extracting formal metrics that relate to cognitive notions of conceptual flexibility and stability.

Our concept network models capture the particular conceptual properties that are associated with an individual concept along with those properties' concept-specific covariation statistics. Individual concept networks are distinct in that properties relate to each other in different ways across basic-level concepts. For example, BLACK and SOFT may co-vary with each other in BANANA, but BLACK and FIRM may co-vary with each other in CHOCOLATE. We found that our network models could successfully discriminate between new conceptual exemplars, suggesting that these within-concept statistics differ reliably between basic-level concepts. These results emerged out of a classification analysis based on eigendecomposition of our concept networks. Eigendecomposition of graphs has previously been used to assess the correspondences between anatomical brain network structure and patterns of functional activation (Medaglia et al., 2017); here we adapted this method to assess the correspondences between conceptual structure and feature-patterns for individual conceptual exemplars. We found that concept networks can simultaneously encode multi-property relationships (i.e. within-concept statistics) and strong single-property contributions, suggesting ways in which information might be organized within the conceptual system and also establishing an exciting new direction for further investigation.

A model structured using within-concept statistics provides a framework in which varied yet appropriate instantiations of a concept may be flexibly activated. An APPLE network may contain a strong connection between CRUNCHY + FRESH and between SOFT + BAKED, enabling the conceptual system to know what sets of properties should be activated in a particular APPLE instance — for example, in the representations evoked by "apple picking" versus "apple pie." The property-covariation statistics for a given concept will determine which sets of properties tend to be co-activated, and how individual properties relate to those sets and to each other. We thus sought to use our compositional concept network models, which contain within-concept statistics, to extract quantitative measures of these phenomena. We found that mean-diversity and core-periphery structure can be interpreted as measures of conceptual flexibility and stability, respectively: a concept network-model's mean-diversity positively predicts semantic diversity (SemD; Hoffman et al., 2013), a network-model's core-periphery fit positively predicts semantic stability (mean cosine similarity; Hoffman et al., 2013), and these two network measures are negatively related to each other across our concepts. Our results also suggest that traditional property-vector models are better at capturing the representation of stable versus flexible concepts, suggesting that a different kind of conceptual model may be necessary to capture the intrinsic flexibility of the conceptual system. We argue that a network-based model of basic-level concepts is one such option.

Network mean-diversity was extracted by averaging over individual nodes' diversity coefficients, a version of the participation coefficient calculated using Shannon entropy. In network neuroscience, participation coefficients have been related to the flexibility of functional network hubs. As discussed above, networks can be partitioned into communities of modules, and nodes can differentially participate in these modules: nodes that only have connections to one module have low participation coefficients, whereas nodes that have connections to many modules have high participation coefficients. This measure is used to differentiate between different kinds of network hubs. High-degree nodes that have low participation coefficients are classified as "provincial" hubs, since they are only connected to one local module. High-degree nodes that have high participation coefficients are classified as "global" or "connector" hubs, since they have connections to many modules across the network (van den Heuvel, 2013; Power et al., 2013). Empirical studies have revealed that the fronto-parietal and cingulo-opercular networks contain high-participation nodes, suggesting that these regions contain functional network hubs (Sporns, 2014). It is further argued that "flexible network hubs" are characterized by their ability to connect with a diverse set of brain regions (Sporns, 2014).

Networks have varying degrees of flexibility with respect to how their states can shift with time. On a higher level, networks can differ in how their underlying *structure* changes with time. In order to study the fluctuations in such dynamic neural networks, Bassett et al. (2011) quantify changes in community structure over time. Instead of extracting a measure from a single network, they define flexibility by analyzing multiple, consecutive networks and determining whether individual nodes either maintain or switch allegiances between communities. Given that networks can exhibit flexibility at these different levels of analysis, we might expect some features of static networks to support flexibility within dynamic networks. We believe that node-participation — and node-diversity, specifically — captures a network characteristic of community allegiance similar to Bassett et al. (2011). Furthermore, high-participation or high-diversity nodes (i.e., global hubs) should play a key role in flexibility because their dense, variable connections mediate transitions between network states. We thus argue that diversity might index flexibility across multiple levels of analysis, and that concept network diversity can be reasonably interpreted as a measure of conceptual flexibility. We do not rule out the possibility that there are other candidate network measures that can also capture conceptual flexibility, and intend to explore this moving forward.

Other frameworks have the potential to capture the flexibility of the conceptual system; these include attractor networks (e.g., Cree et al., 1999; 2006; Rodd et al., 2004) and recent updates of the hub-and-spoke model (Lambon Ralph et al., 2016). The concept network framework proposed here is not in opposition with these other approaches; the development and implementation of all of these methods will greatly benefit our understanding of the semantic system. However, we do believe that a network science approach to conceptual knowledge has its unique advantages. Most broadly, the vast network science methodological toolkit allows us to translate our analyses between cognitive and neural levels of analysis. Given a sample of concept networks and extracted measures of interest, we can observe the functional neural networks recruited for conceptual processing and explore any correspondences between networks across levels. Network neuroscientists have previously forged links to cognitive processes such as motor-sequence learning (Bassett et al., 2011) and cognitive control (Medaglia et al., 2018), setting a precedent for the application of these methods to cognitive science.

The intersection of network science with control theory is another direction that may prove useful for our current purposes. Network controllability refers to the ability to move a network into different network states, and has been applied to structural brain networks in order to shed insights into how the brain may guide itself into easy- and difficult-to-reach functional states (Gu et al., 2015). There have been additional attempts to link brain network controllability to cognitive control (Medaglia, 2018). The application of control theory to concept networks may provide an additional way to quantify conceptual flexibility by identifying nodes that are well-positioned to drive the brain into diverse, specific, or integrated states. Perhaps concept networks that are more controllable overall — that is, networks in which it is easier to reach varied network states — correspond to concepts that are more cognitively flexible.

The concept network framework permits the application of spreading activation models to assess how information flows through these networks. De Deyne et al. (2016) paired a spreading activation model over relational semantic networks and found that their network models could make accurate predictions regarding semantic similarity judgments. In our case, we could use spreading activation models to observe patterns of network activity as a result of different inputs (i.e., contexts). For example, we can attempt to model how conceptual information differs when presented in an adjective-noun phrase: a vector representing adjectival information could be provided as input to the networks, resulting in a specific pattern of activity across the nodes. More generally, these techniques might provide a way to model context-dependent conceptual meaning.

We acknowledge that our proposed concept network model framework has some limitations. First, a large amount of data needs to be collected to enable the calculation of within-concept statistics. However, now that we have established the feasibility and usefulness of these models, it is possible to develop online platforms that can streamline data collection. Second, there is a sense in which some of the methodological decisions are arbitrary: for example, the number of properties to model as network nodes, and the number of concept-states, could influence the resulting networks and extracted measures. These concerns are mitigated, however, by our findings that results calculated from our concept models in Set 1 and Set 2 were comparable, and that the number of concept-states did not significantly predict our network measures of interest. Using standard network filtering methods also reduces these concerns, since it decreases the number of arbitrary parameter decisions required by the experimenter. It is also imperative that the measures extracted from concept networks are interpretable in the context of conceptual knowledge. Only by understanding the potential correspondences between network structure and the structure of conceptual knowledge will these ideas prove useful.

Here we have constructed concept network models, confirmed their ability to capture concept-specific information, and extracted network measures that relate to cognitive measures of conceptual flexibility and stability. We believe the application of network science to conceptual knowledge will provide a set of tools that will enable the intrinsic flexibility of the conceptual system to be explored and quantified.

# Acknowledgements

This work was supported by an NSF graduate research fellowship awarded to SHS, DP5-OD021352 awarded to JDM, and National Institute of Health grant R01 DC015359 awarded to STS.

# Appendix

### Concept-States (Set 1)

CHOCOLATE: bittersweet chocolate, caramel chocolate, chocolate bar, chocolate chips, chocolate syrup, cocoa powder, dark chocolate, chocolate fudge, melted chocolate, milk chocolate, nut chocolate, salted chocolate, white chocolate, hot chocolate

BANANA: banana chips, banana pudding, Cavendish banana, fried banana, frozen banana, mashed banana, over-ripe banana, peeled banana, plantain, raw banana, red banana, rotten banana, sliced banana, unripe banana, ripe banana

BOTTLE: baby bottle, beer bottle, broken bottle, juice bottle, liquor bottle, medicine bottle, milk bottle, soda bottle, spray bottle, water bottle, wine bottle

TABLE: bedside table, changing table, coffee table, conference table, dining table, drafting table, end table, folding table, kitchen table, play table, poker table, pool table, side table, workbench

PAPER: butcher paper, cardboard, cardstock, construction paper, envelope, graph paper, legal paper, newspaper, notebook paper, paper towel, papyrus, poster board, printer paper, sandpaper, scrap paper, sketch paper, stationery, tissue paper, toilet paper, wrapping paper, writing paper

### Concept-States (Set 2)

KEY: car key, key to a city, door key, encryption key, garage key, key to my heart, house key, key card, keyboard key, map key, master key, motorcycle key, office key, padlock key, password, piano key, key to a safe, skeleton key

PUMPKIN: pumpkin bar, pumpkin bread, pumpkin candle, canned pumpkin, pumpkin cookie, pumpkin in a field, Halloween pumpkin, Jack-O-Lantern, pumpkin latte, pumpkin muffin, pumpkin pie, pumpkin puree, rotten pumpkin, pumpkin seeds, smashed pumpkin, pumpkin soup, pumpkin spice, whole pumpkin, Thanksgiving pumpkin

GRASS: astroturf, bamboo, barley grass, bent grass, grass clippings, crab grass, dead grass, hay, lawn grass, lemongrass, marijuana, oat grass, overgrown grass, grass seeds, sod, wheatgrass

COOKIE: almond cookie, butter cookie, chocolate cookie, chocolate chip cookie, Christmas cookie, cookie cake, cookie dough, ginger snap, Girl Scout cookie, lemon cookie, M&M cookie, macadamia nut cookie, macaroon, mint cookie, no-bake cookie, oatmeal raisin cookie, Oreo cookie, peanut butter cookie, shortbread, snickerdoodle, sugar cookie, wafer cookie

PICKLE: bread and butter pickles, canned pickles, pickle chips, chopped pickles, cucumber pickles, dill pickles, garlic pickles, gherkins, hamburger pickles, homemade pickles, jarred pickles, kosher pickles, relish, sliced pickles, sandwich pickles, pickle spears, whole pickles

PILLOW: airplane pillow, bed pillow, body pillow, cotton pillow, couch pillow, decorative pillow, down pillow, feather pillow, foam pillow, hypo-allergenic pillow, memory foam pillow, neck pillow, silk pillow, throw pillow, travel pillow

KNIFE: bread knife, butcher knife, butter knife, cheese knife, chef's knife, dagger, hunting knife, jackknife, machete, paring knife, pocket knife, steak knife, switchblade, sword, throwing knife, utility knife

WOOD: wood blocks, wood chips, chopped wood, wood fence, firewood, floor, wood furniture, log, lumber, wood paneling, paper, planks, plywood, wood pulp, sticks, tree, cedar wood, cherry wood, maple wood, oak wood, pine wood, walnut wood

PHONE: android phone, antique phone, broken phone, car phone, cell phone, emergency phone, flip phone, home phone, iPhone, land line, pay phone, rotary phone, satellite phone, smart phone, wall phone, wireless phone

CAR: broken down car, compact car, convertible, coupe, electric car, family car, hatchback, hybrid car, jeep, luxury car, pickup truck, race car, rental car, sedan, sports car, station wagon, SUV, toy car, truck, used car, van